# Lessons from the History of European EMU

**CHRIS KIRRANE**

**Abstract**

This paper examines the history of previous examples of EMU from the viewpoint that state actors make decisions about whether to participate in a monetary union based on rational self-interest concerning costs and benefits to their national economies. Illustrative examples are taken from nineteenth century German, Italian and Japanese attempts at monetary integration with early twentieth century ones from the Latin Monetary Union and the Scandinavian Monetary Union and contemporary ones from the West African Monetary Union and the European Monetary System. Lessons learned from the historical examples will be used to identify issues that could arise with the move towards closer EMU in Europe.

**Introduction**

After the agreement of Maastricht, the European Community now aims to complete full monetary integration. European politicians seem convinced but economists differ in their opinions. Feldstein for example (1992) recently tackled the journey to European Monetary Union (EMU). Others, such as Graue (1992) defended the EMU against counter arguments. So it is necessary to place the current debate in a historical perspective. There are already several tests of monetary integration of currencies inside a state or across national borders. Some met with success, others not. It is instructive to study the beginning and evolution of other monetary unions and to examine the factors for success or failure. This paper will present a short outline of some monetary integration historical experiments to examine if they can offer explanation about the future of monetary integration in Europe.

**The conceptual framework**

Before approaching the description of the historical events, it is useful to clarify certain conceptual elements, the term 'monetary integration' or 'monetary unification' includes various levels of integration.

1) In its weak form, it means joining of national currencies at fixed parities within more or less narrow bands but without communal reserves or a common central bank. It is what Corden (1972) has called pseudo-union of exchange rates. The coordination of the economic policy, particularly of monetary policy, is necessary to prevent balance of payments disequilibria.

2) The level of monetary integration is increased by establishment of public confidence in the irrevocability of fixed exchange rate parities. This confidence generally does not appear until after a long period of transition during which lasting fixed exchange rate parities are maintained successfully or confidence gains ground after some sort of political unification.

3) However, complete monetary integration or unification is only achieved when a common currency issued by a single central bank circulates in the monetary union zone. The fundamental benefit of monetary integration is microeconomic and arises from the possibility of reducing information and transaction costs. Monetary integration can generate more trade as a result of the reduction or even disappearance of the uncertainty caused by exchange rate fluctuations of members' national currencies. This benefit exists in the pseudo-union of exchange rates but is only fully achieved with the appearance of public confidence in the fixed exchange rates. It is only at the time of the complete realisation of monetary unification that savings in transaction costs of currency conversion can be made. These benefits are closely associated with the symbolic functions of money as a means of obtaining goods at a cost less than in bartering as the use of the money reduces information costs. These benefits are non-competing in consumption; the prosperity of one member does not reduce of that of other members. This community in consumption is one of the characteristics of a public good.

Monetary integration also produces a secondary benefit, macroeconomic this time. As Mundell's theory indicates, the effect of real and geographically specific shocks can be absorbed by flexible exchange rates. Recent studies on choices of exchange rate regimes show that monetary shocks can be better controlled with fixed exchange rates than with floating rates (Boyer, 1978, Fukuda and Hamada 1987). But, if the primary benefits of the monetary union are microeconomic, the costs are essentially macroeconomic. Monetary independence of national economies becomes more and more limited, mainly when international mobility of capital is important and when wages and prices are rigid for unspecified reasons. This can lead to the sacrificing of locally desired levels of unemployment and prices. A floating exchange rate regime gives national economies the possibility to follow the minimax strategy in the game of monetary policies. In joining a monetary union, a nation must give up this position and adhere to the mutual consensus which results from policy coordination. As nations differ in their growth rates and their choices between inflation and unemployment, a fixed exchange rate regime means participating nations will have to sacrifice their national policy objectives.

Another cost, which was recognised in the nineteenth century, is the foregoing of the seigniorage incomes. In coinage systems which allowed other currencies to circulate within the interior of their borders as legal tender, the host nations gave up the fiat coinage charge which they have could have benefitted from issuing their own fiat coin currency. Today, seigniorage is paid through holding a currency on which no return is paid in for example banking reserves.

The differences between potential national members of a monetary union and their degree of dependence with regard to the seigniorage incomes means that the opportunity costs of foregoing these incomes are unequally distributed and big differences are even capable of destabilising the union, Grilli (1989). Although certain nations are likely to lose seigniorage on their own currencies by the fact of the creation of a unified currency, seigniorage income may be generated from the common currency but would be dependent on the mechanisms stipulated for the distribution of common incomes between the members.

**Costs and the benefit of monetary integration have several characteristics**

Firstly, contrary to the benefits of monetary integration which are international and have a public good aspect, the sacrifices required by the monetary union are especially national. This contrast between the costs and benefits of monetary integration is a crucial factor in the calculation of participation in monetary integration.

Secondly, the final cost-benefit ratio of the countries taking part in a union varies according to time. In the beginning, the costs of sacrificing national economic objectives and an independent monetary policy are important. The more integration of the capital market progresses, the more the financing of deficits becomes easier and thus the costs of adjustment become smaller. On the other hand, the common benefits of monetary integration are not perceptible until the later phase. For example, the savings on the cost of conversion happens only after complete union of exchange rates has been reached and the benefit coming from the stability of exchange rates can be harvested only once public confidence in the fixed nature of exchange rates has been established. Thus, the benefits can occur only in the long run and their realisation is not certain whereas the costs in the early stage of sacrificing an independent monetary policy are certain.

Thirdly, the opening up of a monetary union participating nation's economy to trade has a significant influence on the importance of the benefits and the costs derived from monetary integration. If a nation is open, with large flows of import/export compared to transactions domestic, the cost of adjustment of its GNP or of its employment level for balance of payments reasons will be weak (McKinnon, 1963). If the nation's economy is closed, the price will be raised. Thus, the integration of the markets for goods can provide sufficient conditions for the feasibility of monetary integration by the increase in benefits and the reduction in costs. However, Feldstein (1992) defends the idea that monetary integration is not necessary to enjoy the benefits of other forms of economic integration.

One cannot however just stick to the theory of costs and benefits to understand the process of monetary integration. It is necessary to analyse the motivations or intentions of each participant, that it is a nation or a region inside a nation, which motivate it to join a monetary union (Hamada, 1985, ch.3). In other words, it is necessary to study the political economy of the creation of a monetary union.

According to the theory of rational participation (for example, Riker and Ordeshook, 1973), an individual decides to take part in a collective action if the benefits expected are larger than the costs. For a nation, a rational decision to join a monetary union is exerted if the benefit that it expects, those which come from the use of a common currency and of the reduction of uncertainty, are more important than the price to pay, for example the abandonment of independent monetary policy.

When the benefit of a collective action presents a characteristic of a public good, the amount can, however, be less than Pareto optimal. Olson found out by applying the theory of public goods to collective action (Olson, 1965; Olson, and Zeckhauser, 1966). According to the principle of measurement of public goods according to Samuelson, a rational individual decides the level of the public goods by balancing the marginal private benefit of a public good with the marginal cost of supplying a unit of this good. Thus, the allocation of the public good can be less than the optimal theoretically because an individual decision does not take into account the external effect on the other decision units. Consequently, even when a consensus exists around the objective of collective action, the benefit it produces by this action can be too weak. Here appears an interesting assumption on group behavior, that the behaviour of a large group would be different from that of small group.

A second assumption, the decision unit, if it receives a relatively large proportion of the benefit of the public goods, will be able to support more than its share of the proportional cost. In other words, if each participant acts rationally according to the calculation of the private cost-benefit ratio, a small decision unit can thus exploit larger ones by having a disproportionate influence (Olsen, 1965). One can also interpret, from this point of view, the result of Casella (1990) concerning the effect of the relative size of the country. The theory of Olsen on collective action can also be supplemented by the theory of political entrepreneurship or leadership studied by Wagner (1966) and developed more in detail by Frolich, Oppenheimer and Young (1971). If an agent with political entrepreneurship can persuade the group of the effectiveness of collective action in spite of the apparent excess of individual cost compared to individual benefit, the right quantity of collective goods can then be provided, with a surplus being available for this agent.

The nation's decision concerning their participation is founded on the comparison of the costs and benefits of joining a union. A new member joining a monetary union allows existing members to gain from the previously mentioned external benefits. But the new potential member only takes into consideration its own individual costs and benefits which may be smaller than optimal as the individual member does not take into account the group externalities of existing members (Casella, 1990; Hamada, 1985).

**Monetary Integration in Nation States**

Examples of monetary integration in the process of forming of new nation states can be found in Italy, Germany and Japan in the nineteenth century. These examples are particularly interesting because the process of monetary unification involved unifying currencies issued by local regions into a single national currency (Hamada, 1985, ch.3). The recent economic, monetary and social union of Germany offers a modern example of this phenomenon.

Germany

The creation of Germany in 1871 was made following a long and progressive period of economic harmonisation. In 1834, the Zollverein was created and comprised of a customs union of eighteen German States which abolished interior barriers to trade (Henderson 1984). The collection of customs duties at the Union's borders and their distribution amongst Member States required a stable means of exchange between those who had their own currency. In 1838, the states of the Zollverein agreed to fix the values of their currencies to the mark of Cologne. Two currencies associated with the bigger states prevailed: the thaler of Prussia was the principal currency of North Germany while the guilder was used in Bavaria and in the other states of the south. Prussia which was in charge of movement for the setting up of the Zouverein thus played the main role in the whole process of monetary unification.

Although this agreement helped to establish a firmer relation between the two major monetary blocs, there remained seven currencies in circulation in the German states at that time. Silver was the effective base of all these currencies. Thirty-three banks had the right to issue bank notes. Monetary unification was done in three steps. First of all, the mark became the new monetary unit in 1871 and was backed by gold. The value of the mark was defined according to the silver coins in use which were gradually withdrawn from circulation. The transition to the gold standard was consolidated during the second phase of unification in 1873 when the use of the silver money was limited to the small divisional coins.

Germany was the second power, after Great Britain, to adopt the gold standard and caused an international dash for this standard as a little later other large nations followed its example. Lastly, in 1875, the Bank of Prussia became the Reichsbank. The other banks kept their right to issue but were gradually circumscribed until 1935 when the last five banks saw this privilege abolished. Monetary unification was largely accomplished by 1875, at the end of the third phase, only four years after political unification but it had followed several years of gradual economic integration.

Italy

The unification of Italy in 1861 was done so quickly that there was little thought given to setting up state structures and institutions. At the time, there were five different currencies and five issuing banks, which rose to six when Rome was added as a region in 1870. It was decided that the Piedmontese lira would become the new national currency, the Italian lira. The lira became the basic unit in a bi-metallic system where the gold-silver conversion rate was fixed at 15.5 to 1. Other coins were to be withdrawn and reminted. The various district banks resisted attempts to withdraw their right of issue. In 1893, however, the Bank of Sardinia, which was growing rapidly, merged with the Tuscan banks to become the Bank of Italy.

Japan

In the Tokugawa period, the central government monopolised the issuing of currency which was founded on a bi-metallic system of gold and silver with the incidental use of copper and iron coins. The right to issue bank notes was left however to the feudal lords of local provinces, subject to the control of central government. It often happened that the local feudal lords came to issue local money to help with the difficulties of their provincial administration such as to provide of liquidity for offsetting the deflationary policies of central government. At the time of the Meiji restoration, 1700 kinds of currencies were issued by 244 provinces, probably the case of the greatest quantity of issuing agents within the same country. After the restoration, the government introduced the yen as the monetary unit with a decimalised system. Coins were struck at the same time out of gold or silver according to a ratio from 1 gold share for 16.11 of silver. Though bi-metallic in theory, the standard was, in fact, initially just in silver (Muhleman, 1985). Local notes were repurchased by the government between 1872 and 1879 and new national banks were awarded the right of issuance. In 1899, after inflation caused by the excessive issuance of non-convertible notes and the deflation which followed, the Bank of Japan became the single bank of issuance. Lastly, Japan came to adopt the gold-standard.

German reunification and the GEMSU

The most recent example of monetary union following political unification is the treaty of German Economic, Monetary and Social Union Treaty signed by West and East Germany which took effect on 1 July 1990. Aided by the need for symbolic political unity but also by the fear of economic instability in the GDR, the treaty involved a two-tier conversion system of Ostmarks (OM) into Deutschmarks (DM). Wages, pensions, contracts of credit and personal savings up to 2000 OM were converted at a rate of 1 per 1. This rate greatly exceeded the black market exchange rate before the union. Others accounts were converted at the rate of 2 OM for 1 DM, giving an average rate of conversion of 1.8 for 1 for all exchanges. West Germany as the partner dominating the union, took the driving role by proposing this union and by guaranteeing some of the tax consequences.

Although it is too early to have scientific research on the economic consequences of this union, Alerlof and others (1991) have already produced some analyses. The conversion of labour costs at preferential level has caused a compression of the cost-price ratio involved for East German firms due to the incapacity to raise their prices as demand for their goods decreased dramatically when East Germans, profiting from their new purchasing power, acquired better quality Western products. Fears of an inflationary spending push by East Germans remained baseless since the rate of the savings rose after unification. But at the same time, costs of harmonisation of the infrastructure and of the social systems, and of the closing or liquidation of companies caught by the tightening of the cost-price ratio, have caused an increase in fiscal debt. For the first time for a long time, a swelling of the budget deficit has caused German rates of inflation to be high when compared to other countries in northern Europe.

**Monetary unions between Nation States**

Monetary unions between states without the intention of politic integration are relatively rare. Three historical examples will be examined: on the one hand, the short-lived monetary union between Germany-Austria, and on the other the decades lasting Latin Monetary Union and the Scandinavian Monetary Union. Then two current examples, the Monetary Union of West Africa and the European Monetary System (and stages towards the European Monetary Union) will be described.

The Monetary Union of Germany

In 1857, to extend the Zollverein, the Monetary Convention of Germany was established between the German member states of the Zollverein and Austria. A new basic currency unit, the Zollpfund was adopted, to replace the Cologne mark and the thaler and the guilder were reassigned new parities. Provisions were made for striking new gold coins, 'crowns', and silver coins, the 'union thalers', which however did not have a wide circulation. The Union with Austria did not last long as the Austrian government quickly rescinded the agreement and preserved its fluctuating paper money. After the Seven Weeks War, the Monetary Convention was formally dissolved in 1867, even though it remained valid between the contracting German states until new measures were agreed after the political unification of 1871. In spite of its short duration, this union contributed to the establishing of monetary relations between the German states which led them to monetary union after their political unification.

Latin Monetary Union

The massive surge in gold circulation, around 1850, after the discoveries of Australian and American deposits, caused a panic in some European countries as gold became cheaper. By 1860, Switzerland had devalued its currency. Similar difficulties in France and Belgium led these three countries to consider a common reaction and resulted in the Latin Monetary Union Convention of 1865, which also included Italy. The Union fixed the proportion of gold and silver in coins, which then became legal tender in the four signatory countries. In 1868, Romania, Greece and the Holy See joined but the latter withdrew after only one year.

The bi-metallism of the Union came under renewed pressure after the discovery of silver deposits in Nevada in 1873. Also the new German state and then Holland (in 1873) chose the gold standard so silver flowed out of these countries into the mints of the Union, with a corresponding backward flow of gold. In 1874, the issuance of the 5 franc silver coin of was limited and in 1878 it was suspended. The Union was then described as being founded on a 'lame standard' with gold as the effective base and silver for additional use (Clough, 1952). The Treaty of the Union was thus amended so that each state would refund out of gold the silver in circulation to the other states on termination of the Union. As France's economic power was higher than the others, the Union was in disequilibrium. The Bank of France accumulated a significant amount of foreign silver coins which according to Willis, on the assumption of liquidation of the Union in 1905, France should have required 250 million in gold from Belgium 270 million from Italy and 14 million from Greece. None of these countries could even have honoured their obligations even when they were reduced by an amendment of the Treaty in 1885. Willis concluded in 1905 that the Union 'was condemned to exist under its present conditions for an indefinite period', in spite of the frequent reports which preached its dissolution. The Union ceased functioning in practice at the time of the declaration of the First World War, when a paper standard was introduced. It was formally dissolved only when Belgium withdrew in 1925.

Scandinavian Monetary Union

Sweden and Denmark formed a monetary union in 1873, to be joined by Norway in 1875. A common currency, the Krone based on gold, circulated in the member countries. In Sweden, the issuing of notes was in hands of the Bank of Sweden and private banks, while in Norway and in Denmark issuance was restricted to the Bank of Norway and at the Bank of Denmark. The union extended to cover circulation of the notes when in 1894 issuing banks of Sweden and of Norway agreed to accept their respective notes at parity. Denmark joined them in 1900. Starting from 1905, the conditions of circulation of the issued notes were amended to allow commissions to be charged on foreign notes. In spite of this additional cost, joint circulation of notes continued until the First World War when the refunding of bank notes was suspended and the Union effectively terminated (Nielsen, 1933).

Contemporary unions

West African Monetary Union

The most durable modern monetary union is the West African Monetary Union, which consists of seven French-speaking countries of West Africa. This Union was created around the Central Bank of the States of West Africa (the BCEAO) in 1962 and continues since then, after an important revision of its operations in 1974. The BCEAO issues a common currency, the CFA franc at a rate of 50 to 1 French franc, with the support of the French government. This support takes today the form of an overdraft facility at the French Treasury where most of the Union's foreign currency reserves are held. In return, France exerts a direct influence on the business of the Union by appointing Board members. France also exerts considerable indirect influence on members through assistance and trade links. However, without the French support to maintain the convertibility of the CFA franc, one of main reasons for the union would disappear.

In the Union, the Ivory Coast, which provides half of the local production, is the dominant nation. Monetary policy inside the Union operates in a very decentralised manner as National Monetary Committees within each member state decide on the issuance of currency within the credit allocation directives decided by the board of BCEAO. The Bank sets rediscounting ceilings, ratios of reserves and the discount rate for each member. The total loan of the bank to a Member State cannot exceed 20% of overall incomes of the previous year. The Union was shaken during these last years by the overvaluation of CFA franc compared to the other currencies, because of its indexation to the French franc.

EMS and EMU

The development of the European Monetary System since its unveiling in 1979 offers an extraordinary contemporary example of the process of monetary integration. The EMS was imagined as a 'symmetrical, flexible version of the Bretton- Woods system' (Keenen, 1992), following old plans of monetary union in Europe. The Werner report of 1970 considered the monetary union for 1980, in spite of the turbulent environment of the Seventies which was not very favourable to economic harmonisation. One understands the process of harmonisation as the tightening of the differences between key indicators of the national economy such as the rates of inflation, interest and unemployment. Intentions with regard to the advantages of a narrower union really developed during the 1980s between the eight original members of the EMS and later joined by Spain in 1989 and the United Kingdom in 1990. After frequent adjustments at the beginning, no major realignments have occurred since 1987.

Gradually, the German Bundesbank was used as the monetary anchor for EMS because of the size of the German economy and the power of the Deutschmark, which was supported by a strong aversion to inflation by the Bundesbank. In a decade when anti-inflationary political commitments were important, the credibility offered to the EMS by Germany was appreciated by politicians in other member counties. Late members to the EMS such as Great Britain hoped that their entry would help them to discipline wages and prices. Adhesion to the EMS clearly meant an anti-inflationary market position while avoiding political blame for the increase in unemployment and the reduction in the incomes in noncompetitive industries. The history of Great Britain since its entry in 1990 did not confirm the fears of its detractors but also did not reinforce the hopes of its enthusiasts: interest rates fell but insufficiently to encourage a recovery from a serious recession.

**Lessons of these historical experiments**

These historical experiments provide several lessons (Hamada, 1985). First of all, political integration invariably preceded complete monetary integration, while economic integration has sometimes preceded and sometimes followed political integration. In Italy however, political integration was done before completion of the free circulation of goods was implemented. In Japan also, free trade was only initiated when the Meiji government completely centralised political power, but free movement of labour arrived only after the Meiji Restoration. The circumstances which precipitated the recent reunification of Germany needed simultaneously economic and political unification.

Secondly, monetary unions across frontiers often did not last long because the political integration necessary for consolidation had not been realised. These unions were effective, at least in the short term, only if the political direction of a dominant state occurred, the number of participants was limited and if broad economic integration had occurred. In the case of the monetary unions which lived a long time, France exerted its leadership of the Latin Monetary Union and France's influence frames the manner in which the Monetary Union of West Africa operates. Domination by Sweden launched the Scandinavian Monetary Union which with only three member countries who had geographic and cultural integration, the union proved to be remarkable stable.

Thirdly, the existence of one or more metal currencies was an effective instrument to maintain confidence in the union's exchange rates. The Latin Monetary Union was founded on a standard of metal currency but the issuing of paper money not backed by metal by Italy and Greece caused serious tensions in the Union. Adhesion to the gold standard helped the Scandinavian Monetary Union. More recently, when currencies lack any metallic base, a will to maintain one strong currency is also used as a basis for a monetary union, as is the case for the WAMU with the indexation to the French franc and for the EMS with its deutschmark leadership.

These discoveries can be related with the calculation of participation. The benefits of the monetary unification only become extensive when strong confidence accompanies the fixing of parities or when one single currency is in circulation. Thus, the metallic content of currencies was important by bringing confidence in the exchange rate union but political integration was necessary to support this confidence over a long period.

Fourthly, these historical experiments illustrate the difficulty involved in maintaining two kinds of currencies in circulation simultaneously as Gresham law remains valid: the currency of better quality was or hoarded or exported, involving a surplus of the currency of lower quality in the union, for example, as in the Latin Monetary Union. In Japan, in the Tokugawa period, a conflict about the metallic content of coins between gold and silver arose between the merchants of Tokyo where gold was more spread and those of Osaka which preferred silver (Oishi, 1974).

Fifthly, the price of the participation in a monetary union in the nineteenth century demanded the sacrifice of seigniorage rather than the cost of policy adjustments as the costs associated with underutilisation of resources to correct a balance of payments problem were hardly felt before the Great Depression (Guggenheim, 1973). Recent studies concerning the EMS have rekindled interest about the seigniorage issues. Emerson and al. (1992) carried out a simulation of the effects of EMU on member countries more dependent on the seigniorial income. According to their estimates, only two (Greece and Portugal) of the four countries accruing more than 1% of their GNP through seigniorage would lose more than 1% of their GNP under EMU. These authors consequently claim that microeconomic welfare benefits gained from the fixing of exchange rate exceeds the occasional costs even for these nations.

**Conclusion**

The movement towards completion of monetary union in Europe which speeded up after the Delors Commission report of 1989, has now slowed. Feldstein (1992) indicated that it was the implicit influence on the EMS by Germany which had pushed other members towards closer monetary integration in the form of the EMU. Members of the EMS ended up approving 'democratised' monetary institutions of EMU rather than informal rule by the Bundesbank. Moreover, recent fears of an increase in German inflation after the Treaty of Reunification has led to fears among members of the inflationary consequences for their economies and they wish to reduce Germany influence.

In Maastricht, a revised process of three stages was adopted. The first stage of liberalisation of capital is already largely completed. The second will start in January 1994, when the European Monetary Institute will be in place to organise the issuance of a common currency and to facilitate monetary cooperation. The third phase will start in 1999, when exchange rates will be irrevocably fixed and effective monetary authority will be allotted to a new European Central Bank. Although the terms of the Treaty of Maastricht must still be approved by all the member countries, a true monetary union is foreseen before the end of this century.

One hundred years ago, Europe already knew a relatively high degree of monetary integration. The appearance of the states of Germany and Italy had already considerably reduced the number of currencies in circulation. The Latin Monetary Union system in France, Belgium Italy, Switzerland and Greece was largely followed by non-member states like Spain, Austria and Finland, thus bringing stability to European monetary exchanges. The Scandinavian Monetary Union eliminated the need for monetary exchange in Northern Europe. The movement towards monetary integration reached its apogee in 1867 with the International Monetary Convention when France sought to convince the other states to join the new Latin Monetary Union. For political and economic reasons, important powers like Great Britain, Germany and the United States refused. The responsibility of maintaining the Union during the following years weighed heavily on France. A century later, this lesson seems to have been learned as France was most insistent on German participation in order to ensure a broader economic base and thus get greater creditability for the process.

There is not a doubt that a move towards monetary union still poses many questions. As the process of political integration in Europe progresses, politicians will seek to create the necessary structures for economic and monetary integration. It remains to be seen if these news structures will be adequate to solve the complex problems of coordination that greater monetary unification will pose.